\documentclass[a4paper,11pt]{article}
\pdfoutput=1 

\usepackage{jinstpub} 
\usepackage{subfigure}
\usepackage{subfigmat}
\usepackage{changepage}
\usepackage{textcomp}
\usepackage{xcolor}
\usepackage{float}
\usepackage{upgreek}
\usepackage{hyphenat}

\newcommand{\micron}{\textmu\rm{m}}
\newcommand{\Neq}{\ensuremath{\,\rm{n}_{\rm{eq}}/cm^2}}

\title{Testbeam results of irradiated ams H18 HV-CMOS pixel sensor prototypes}

\author[a]{M.~Benoit,}
\author[b]{S.~Braccini,}
\author[c]{G.~Casse,}
\author[d]{H.~Chen,}
\author[d]{K.~Chen,}
\author[a,1]{F.A.~Di Bello,}
\author[a]{D.~Ferrere,}
\author[a]{T.~Golling,}
\author[a]{S.~Gonzalez-Sevilla,}
\author[a]{G.~Iacobucci,}
\author[a]{M.~Kiehn,}
\author[d]{F.~Lanni,}
\author[d,e]{H.~Liu,}
\author[a,c]{L.~Meng,}
\author[b]{C.~Merlassino,}
\author[b]{A.~Miucci,}
\author[f,1]{D.~Muenstermann,\note{Corresponding author.}}
\author[a,g]{M.~Nessi,}
\author[h]{H.~Okawa,}
\author[i]{I.~Peri\'c,}
\author[b]{M.~Rimoldi,}
\author[a,g]{B.~Risti\'c,}
\author[a]{M.~Vicente Barrero Pinto,}
\author[c]{J.~Vossebeld,}
\author[b]{M.~Weber,}
\author[b]{T.~Weston,}
\author[d]{W.~Wu,}
\author[d]{L.~Xu}
\author[a]{and E.~Zaffaroni}

\affiliation[a]{D\'epartement de Physique Nucl\'eaire et Corpusculaire (DPNC), University of Geneva, 24 quai Ernest Ansermet 1211 Gen\`eve 4, Switzerland}
\affiliation[b]{Albert Einstein Center for Fundamental Physics and Laboratory for High Energy Physics, University of Bern, Siedlerstrasse 5, CH-3012 Bern, Switzerland}
\affiliation[c]{Department of Physics, University of Liverpool, The Oliver Lodge Laboratory, Liverpool L69 7ZE, UK}
\affiliation[d]{Brookhaven National Laboratory (BNL), P.O. Box 5000, Upton, NY 11973-5000, USA}
\affiliation[e]{Dept. of Modern Physics, University of Science and Technology of China, Hefei, Anhui 230026, China}
\affiliation[f]{Lancaster University, Physics Department, Lancaster, LA1 4YB, UK}
\affiliation[g]{European Organization for Nuclear Research (CERN), 385 route de Meyrin, 1217 Meyrin, Switzerland}
\affiliation[h]{Faculty of Pure and Applied Sciences, and CiRfSE, University of Tsukuba, Tsukuba 305-8571, Japan}
\affiliation[i]{Karlsruhe Institute of Technology (KIT), IPE, 76021 Karlsruhe, Germany}

\emailAdd{Daniel.Muenstermann@cern.ch}

\abstract{HV-CMOS pixel sensors are a promising option for the tracker upgrade of the ATLAS experiment at the LHC, as well as for other future tracking applications in which large areas are to be instrumented with radiation-tolerant silicon pixel sensors. We present results of testbeam characterisations of the $4^{\mathrm{th}}$ generation of Capacitively Coupled Pixel Detectors (CCPDv4) produced with the ams H18 HV-CMOS process that have been irradiated with different particles (reactor neutrons and 18 MeV protons) to fluences between $1\times 10^{14}$ and $5\times 10^{15} \,\textrm{1-MeV-}\Neq$. The sensors were glued to ATLAS FE-I4 pixel readout chips and measured at the CERN SPS H8 beamline using the FE-I4 beam telescope. Results for all fluences are very encouraging with all hit efficiencies being better than 97\% for bias voltages of $85\,$V.  The sample irradiated to a fluence of $1\times 10^{15}\Neq$ -- a relevant value for a large volume of the upgraded tracker -- exhibited 99.7\% average hit efficiency. The results give strong evidence for the radiation tolerance of HV-CMOS sensors and their suitability as sensors for the experimental HL-LHC upgrades and future large-area silicon-based tracking detectors in high-radiation environments.
}

\keywords{Solid state detectors, Radiation-hard detectors, Particle tracking detectors, Electronic detector readout concepts (solid-state)}

\begin{document}

\maketitle

\flushbottom

\section{Introduction}\label{sec:introduction}
To extend the physics reach of the Large Hadron Collider (LHC), upgrades are planned to increase its luminosity allowing for up to 3000 fb$^{-1}$ of data to be collected by ATLAS and CMS. The increase in radiation damage associated with this also requires upgrades to the experiments, in particular the replacement of the Inner Trackers, which requires very large areas of extremely radiation-tolerant silicon detectors \cite{ATLAS-Upgrade-LoI, HL-LHC-paper}. 

Due to the large areas to be instrumented, special care has to be taken to investigate cost-efficient, but still very radiation-tolerant sensor options. Thanks to their large production output, CMOS foundries are capable of producing large areas of silicon at -- compared to bespoke hybrid pixel detectors -- affordable cost. Some CMOS processes developed for the automotive industry (HV-CMOS\footnote{High-Voltage CMOS}) or imaging applications (HR-CMOS\footnote{High-Resistivity CMOS}) are promising candidates for sensor production, thanks to their tolerance to high bulk bias voltages which are necessary for fast charge collection (see e.g. \cite{IvanPeric}). In this work, the radiation tolerance of a test chip produced with the ams H18\footnote{ams {\bf H18} is an {\bf H}V-CMOS process  with {\bf18}0 nm feature size by ams/IBM.} High-Voltages CMOS process \cite{ams} was investigated after irradiation with mixed spectrum reactor neutrons at the TRIGA reactor \cite{Ljubljana} of the JSI, Ljubljana, Slovenia, and with 18 MeV protons at the Bern Cyclotron Laboratory \cite{Bern}, Switzerland.

\begin{figure}[h]
\subfigure[]{\includegraphics[width=.68\textwidth]{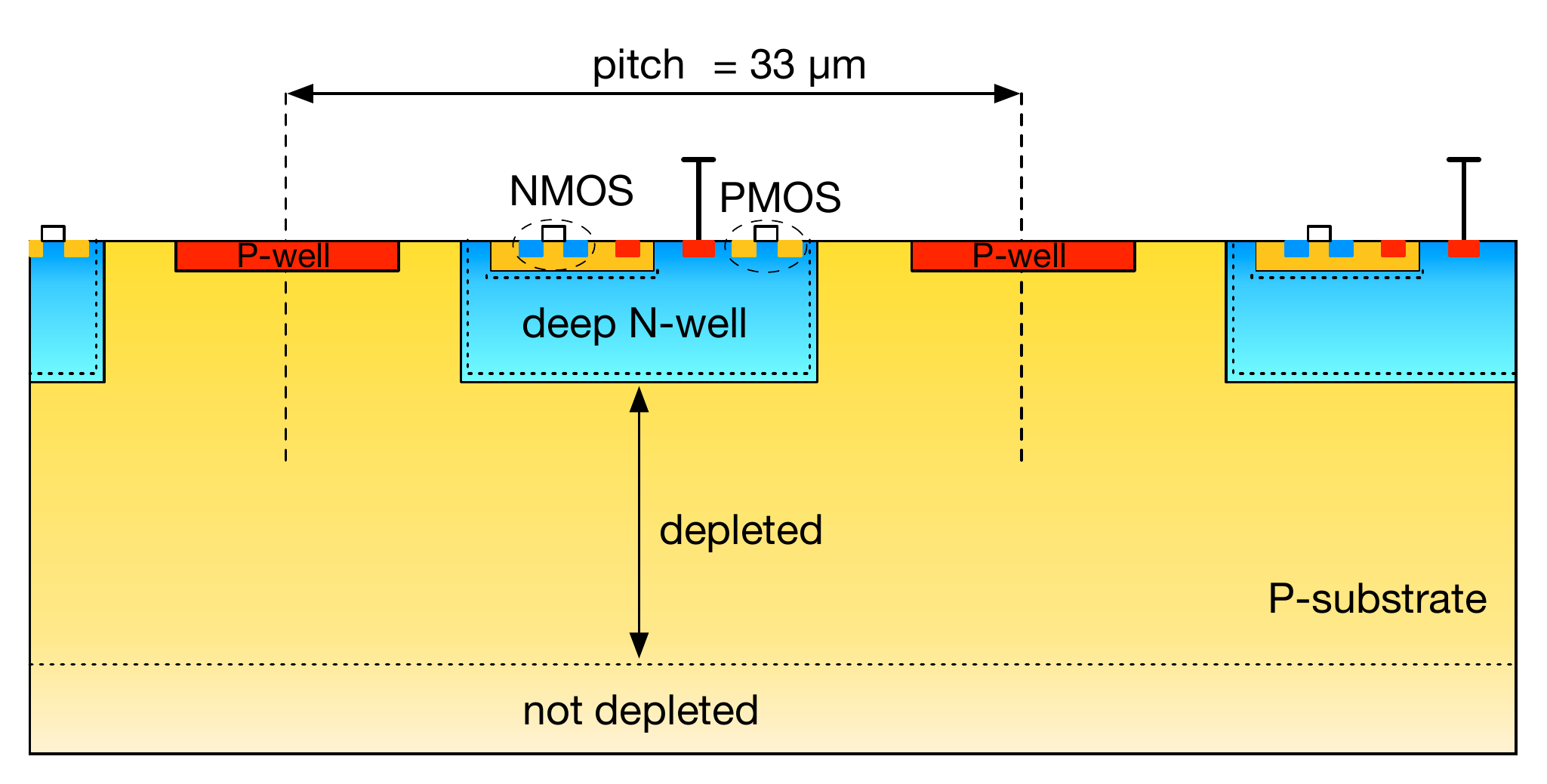}}
\subfigure[]{\includegraphics[width=.3\textwidth]{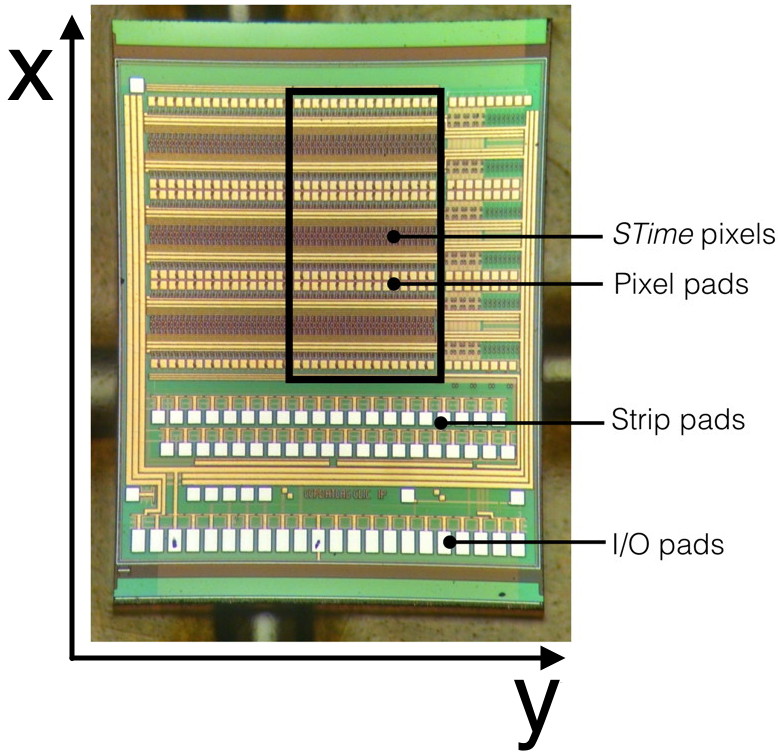}}
\caption{(a) Schematic cross section of an HV-CMOS sensor: the deep n-well is the charge-collecting electrode and also contains additional CMOS circuits such as a preamplifier. (b) Photograph of the CCDPv4 sensor used in this study with the sub-matrix of STime-type pixels marked.}

\label{Sketch_HV_CMOS}
\end{figure}

An important feature of processes suitable for sensor production is the existence of a so-called deep n-well (DNW) in a moderately p-doped bulk, which is necessary to insulate the circuits from high voltages (see figure \ref{Sketch_HV_CMOS}a). The depth of the DNW is a few $\upmu$m, preventing the depletion region inside the n-well from reaching the in-pixel CMOS FETs, which would cause their behaviour to change. The DNW itself must also have a suitable doping profile that avoids too high electric fields at its edges which would cause impact ionisation and thus breakdown.

In its most simple form, a matrix of DNWs could be used as a classical planar n$^+$-in-p pixel detector. However, the nominal resistivity of the standard base material is only 10--20 $\Omega\,\cdot\,$cm, leading to a calculated depletion depth of only about 10--15 $\upmu$m for bias voltages of 80--150 Volts. Such thin layers of depleted silicon would yield a most probable charge of only about 600--900 electron-hole-pairs for a vertically penetrating minimum ionising particle (MIP), clearly a very challenging value for classical pixel readout chips with usual threshold settings of 1500 electrons or above. 

However, the CMOS production process allows for an in-pixel amplification stage. If combined with small pixels that yield low capacitance values, such in-pixel circuits can be used to amplify the signal to a suitable amplitude for a discriminator or directly for a readout chip. Examples for monolithic HV-CMOS sensors produced in the ams H18 process are the MuPix family of sensor prototypes \cite{MuPix}, conceived for the Mu3e experiment at PSI. Sensors relying on dedicated readout chips are e.g. the HV2FEI4/CCPD (\emph{HV-CMOS-to-FE-I4} or alternatively \emph{Capacitively Coupled Pixel Detector}) prototypes \cite{CCPD} intended for use in the upgraded ATLAS detector at the HL-LHC. A schematic cross section of an HV-CMOS sensor is depicted in figure \ref{Sketch_HV_CMOS}.

CCPD sensors are designed to be coupled to a pixel readout chip. As suggested by the name, this can be done by capacitive coupling using non-conductive glue since the amplifier and discriminator output signal is large and fast enough to not be affected by the glue-filled gap in the signal path. For comparison, the classical bump-bonding process is also possible. For R\&D purposes, the sensor can also be operated standalone, and pixel outputs from the matrix can be investigated by configuring internal multiplexers. In this way, however, only individual pixels can be observed. In addition, several test structures and circuits have been added to the prototypes. For this study, ATLAS FE-I4 \cite{FE-I4} IBL\footnote{Insertable b-layer, the new innermost 4$^{\textrm{th}}$ pixel layer of ATLAS at a radius of only 3.3 cm.} \cite{ibl} readout chips  were glued to CCPDv4 HV-CMOS sensors, which was the latest generation of capacitively coupled HV-CMOS sensors available at the time of the beam test.

\section{HV-CMOS devices under test}
\subsection{CCPDv4}
\begin{figure}[!htbp]
\subfigure[]{\includegraphics[width=0.6\textwidth]{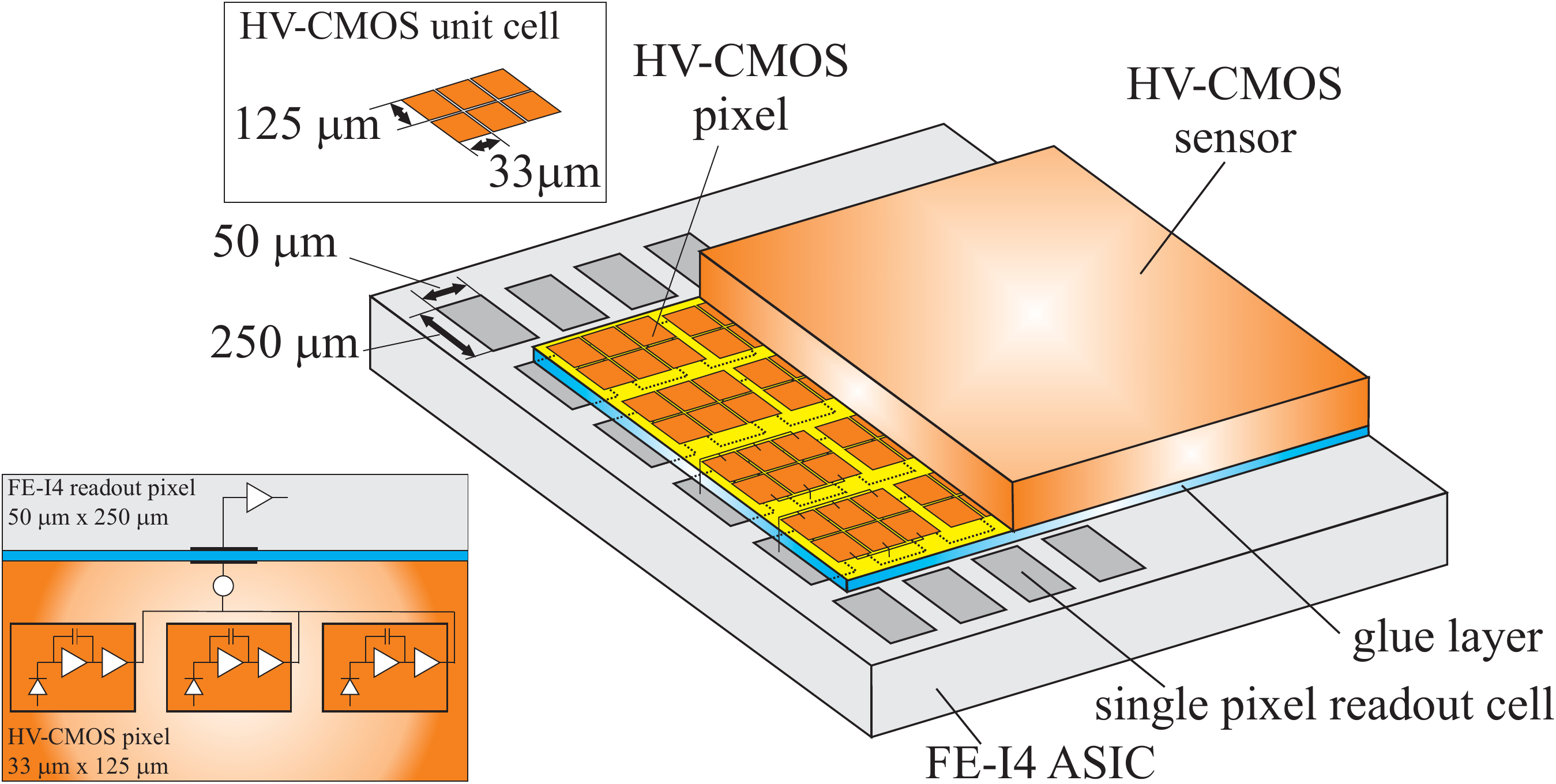}}
\subfigure[]{\includegraphics[width=0.4\textwidth]{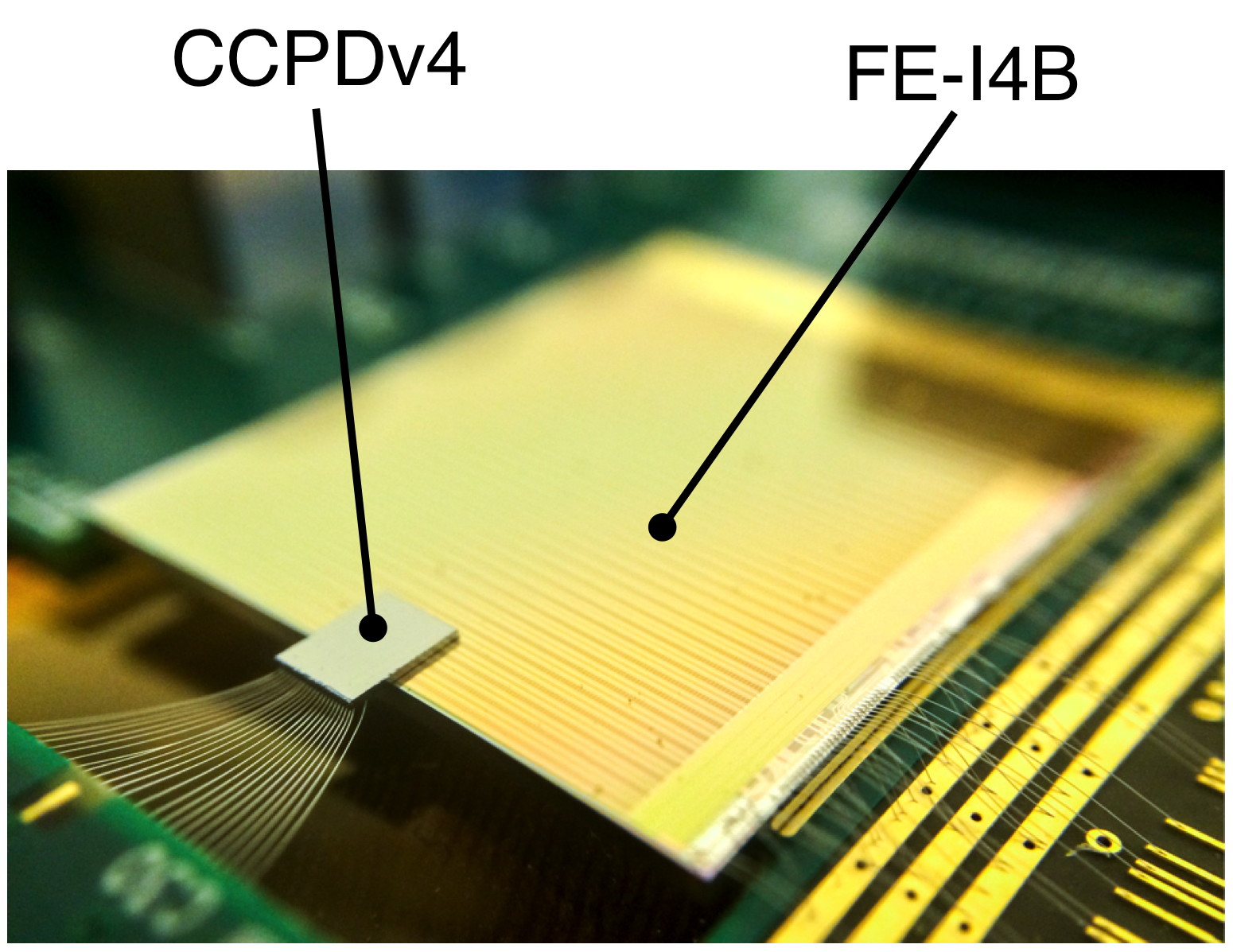}}
\caption{(a) Schematic representation (not in scale) of the HV-CMOS to FE-I4 connections. The bottom-left inset shows how three HV-CMOS pixels (forming a so-called {\em macro-pixel}) are capacitively coupled to a single FE-I4 readout pixel. (b) Final assembly of a FE-I4 pixel readout chip to a HV-CMOS CCPDv4 sensor via capacitive coupling. \cite{CCPDv4-paper}}
\label{fig:dut}   
\end{figure}

CCPDv4\footnote{Sometimes also referred to as \textbf{AMS180v4}} sensors -- the 4$^{\textrm{th}}$ generation of test sensors produced in the ams H18 process using regularly scheduled multi-project wafer productions -- were produced on a nominal 10~$\Omega\,\cdot\,$cm substrate; details about the creation of the samples can be found in \cite{CCPDv4-paper}, therefore only a brief summary will be presented here. The design rule set specifies a minimum breakdown voltage of 60\,V, the experimental breakdown voltage was found to be at 93\,V. However, the noise gradually increases already at bias voltages above 80~V. The CCPDv4 contains a pixel matrix matching the FE-I4 with several different pixel flavours, out of which this study focused on the performance of the so-called \emph{STime}-type pixels (see figure \ref{Sketch_HV_CMOS}b). Each \emph{STime} pixel on the HV-CMOS sensor features a size of only 33 \micron\ by 125 \micron\ and contains an amplifier and a discriminator together with a 4-bit in-pixel Tune-DAC, allowing a per-pixel threshold tuning. To match the FE-I4 pixel size of 50 \micron\ by 250 \micron, the HV-CMOS pixels are grouped together in gangs of three, forming \emph{two} 100 \micron\ by 125 \micron\ large macro-pixels for each \emph{two} neighbouring 50 \micron\ by 250 \micron\ pixels (see figure \ref{fig:dut}a). While the chip allows the encoding of the hit pixel's coordinate, this feature was not used during this study and testbeam reconstruction was performed on the macro-pixel level. However, the threshold tuning of the HV-CMOS matrix was performed on the pixel level. The CCPDv4 sensors were glued to FE-I4 readout chips (see figure \ref{fig:dut}b) using Araldite 2011 non-conductive epoxy glue on a SET Acc$\upmu$ra~100 flip-chip bonder at the University of Geneva with a typical glue thickness of less than 1 $\upmu$m. The calibration factor between deposited charge and threshold voltage of ${\sim}8.6$ e$^-$/mV from \cite{CCPDv4-paper} can also be applied to this work, however, as it has not been re-measured, the thresholds are given in mV throughout this work.

\subsection{Irradiated samples}

In this paper, fluences are given in 1-MeV-Neutron equivalents per area, abbreviated to $\Neq$. 

The relevant fluences for the application in the upgraded ATLAS Inner Tracker (ITk) at HL-LHC (target integrated luminosity of 3000 fb$^{-1}$) range between \emph{about} $2\times10^{14}$ (outer strip layers), $1\times10^{15}$ (outermost pixel layer), $4\times10^{15}$ (pixel layer 1, assuming exchange after 1500 fb$^{-1}$) and $1\times10^{16}$  \Neq\ (pixel layer 0, assuming exchange after 1500 fb$^{-1}$) \cite{PaulMiyagawa}. The most relevant range for the ATLAS HV-CMOS pixel demonstrator project is in the region of the outermost pixel layer, i.e. around $1\times10^{15}$ \Neq, where the area to be instrumented is comparatively large.

Earlier measurements \cite{CHESS1, marcos} suggested that the fluence region with the worst collected charge/hit efficiency is at few $10^{14}$ \Neq . This is attributed to the set-in of trapping, causing the loss of diffusion. The performance recovers at about $1\times10^{15}$ \Neq\ due to an increase of the depletion zone outweighing the reduced diffusion. The latter effect is attributed to a suspected acceptor removal effect. In view of these findings, this work focussed on the fluence region around the recovery point.

While it is very difficult to irradiate samples mounted on a PCB in a reactor, this is possible for irradiation with 18 MeV protons at the Bern Cyclotron \cite{Bern}. Therefore, one sample was subsequently irradiated to $1.3\times10^{14}$  \Neq\  and $5\times10^{14}$  \Neq\ at Bern, while two individual samples were irradiated to $1\times10^{15}$  \Neq\  and one to $5\times10^{15}$  \Neq\  with reactor neutrons at the TRIGA reactor of the JSI in Ljubljana \cite{Ljubljana}.

\section{Testbeam experimental setup}

\begin{figure}[!hbtp]
\begin{center}
\includegraphics[width=0.9\textwidth]{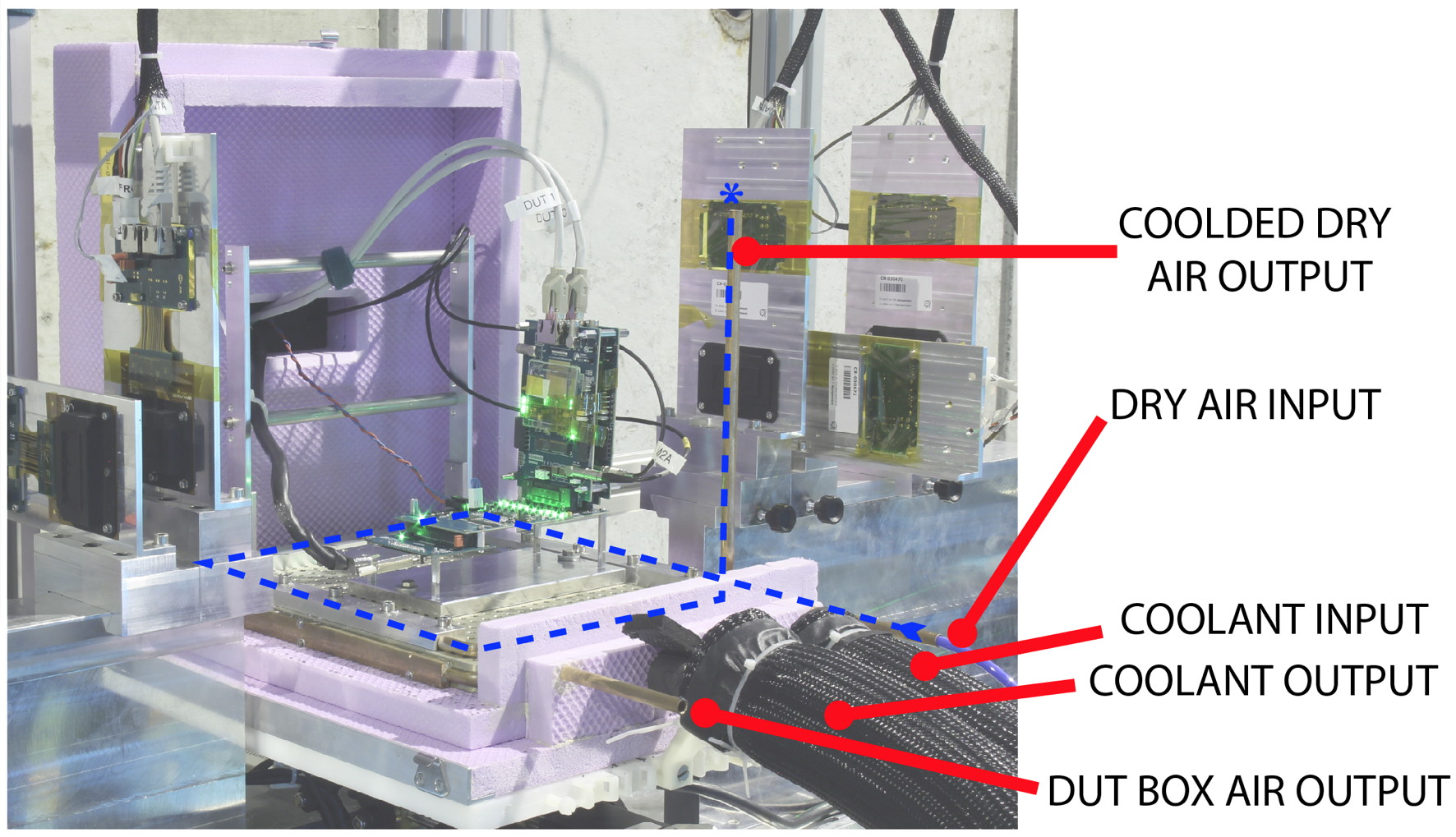}
\caption{Photograph of the opened cooling box. The insulating foam of the remaining side wall containing the feedthrough for the DUT connection cables can be seen in violet. To the left and right, FE-I4 telescope planes are visible. The coolant and nitrogen inputs/outputs as well as the exhaust pipe are highlighted.}
\label{fig:dut-box}
\end{center}
\end{figure}

The main part of the experimental beam test setup has already been described in earlier publications on the FE-I4 telescope itself  \cite{telescopePaper} and also in a paper about testbeam results with an unirradiated CCPDv4 sample \cite{CCPDv4-paper}. 

For this study, cooling of the irradiated devices under test (DUTs) had to be provided to reduce the radiation-induced leakage current. To this end, a dedicated cold box has been implemented (see figure~\ref{fig:dut-box}). It consists of a cooled base plate housed inside an enclosure of a rugged insulating foam. The copper base plate contains a regular array of threaded holes (M5, 1\,cm pitch) to be able to conveniently fix samples inside. Unlike earlier approaches, such as the DOBox \cite[p.~49 ff.]{DissTroska}, the base plate is not cooled by dry ice, but by a Huber Unistat chiller, which circulates a silicon oil coolant (that can be used to cool down to $-75\,^\circ$C) through a copper tube that is glued into in a milled semi-circular groove around the edge of the base plate. On top of the plate, a pipe is attached to the cooled plate to pre-cool the nitrogen used for flushing the box to avoid condensation; this pre-cooling is essential to reach low sample temperatures. The targeted on-sensor temperature for the ITk Pixel Detector -- and therefore also for our cold box to be as realistic as possible -- is $ -25\,^\circ$C.

The final setup can be seen in figure \ref{fig:dut-box}. During the commissioning of the box, the temperature of the CCPD DUT and FE-I4 readout chip reached $-28.6\,^\circ$C and $-26.9\,^\circ$C, respectively, with a nitrogen temperature of $-28\,^\circ$C. Both CCPD and FE-I4 were powered on.  The temperature measurements were done with an uncertainty of $2\,^\circ$C, the measured relative humidity typically is 5--7\%.

\section{Results}
\subsection{Hit Efficiencies}

\begin{figure}[b]
\subfigure[]{\includegraphics[width=.3\textwidth]{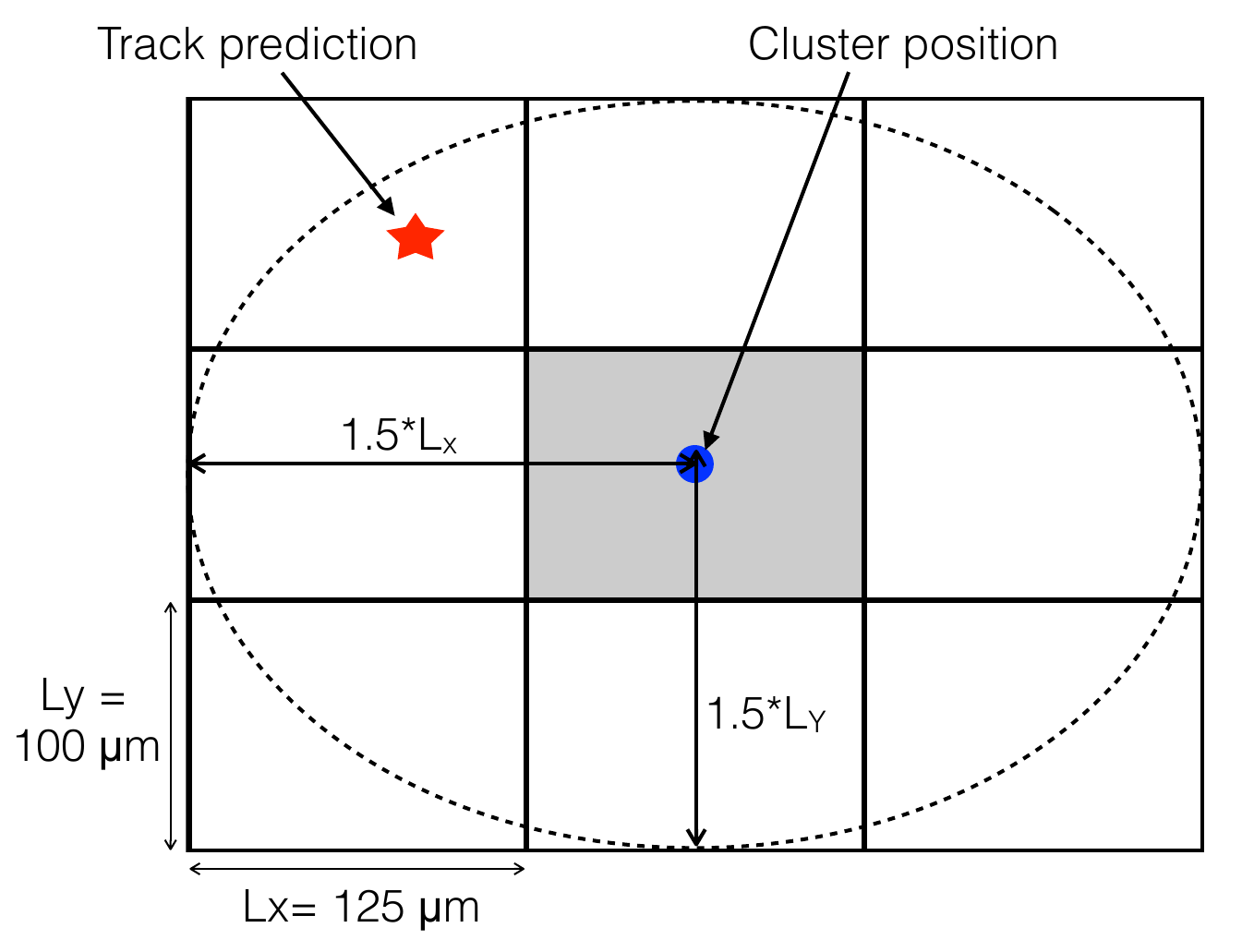}}
\subfigure[]{\includegraphics[width=.34\textwidth, ,trim=0cm 0cm 1.5cm 0.5cm, clip]{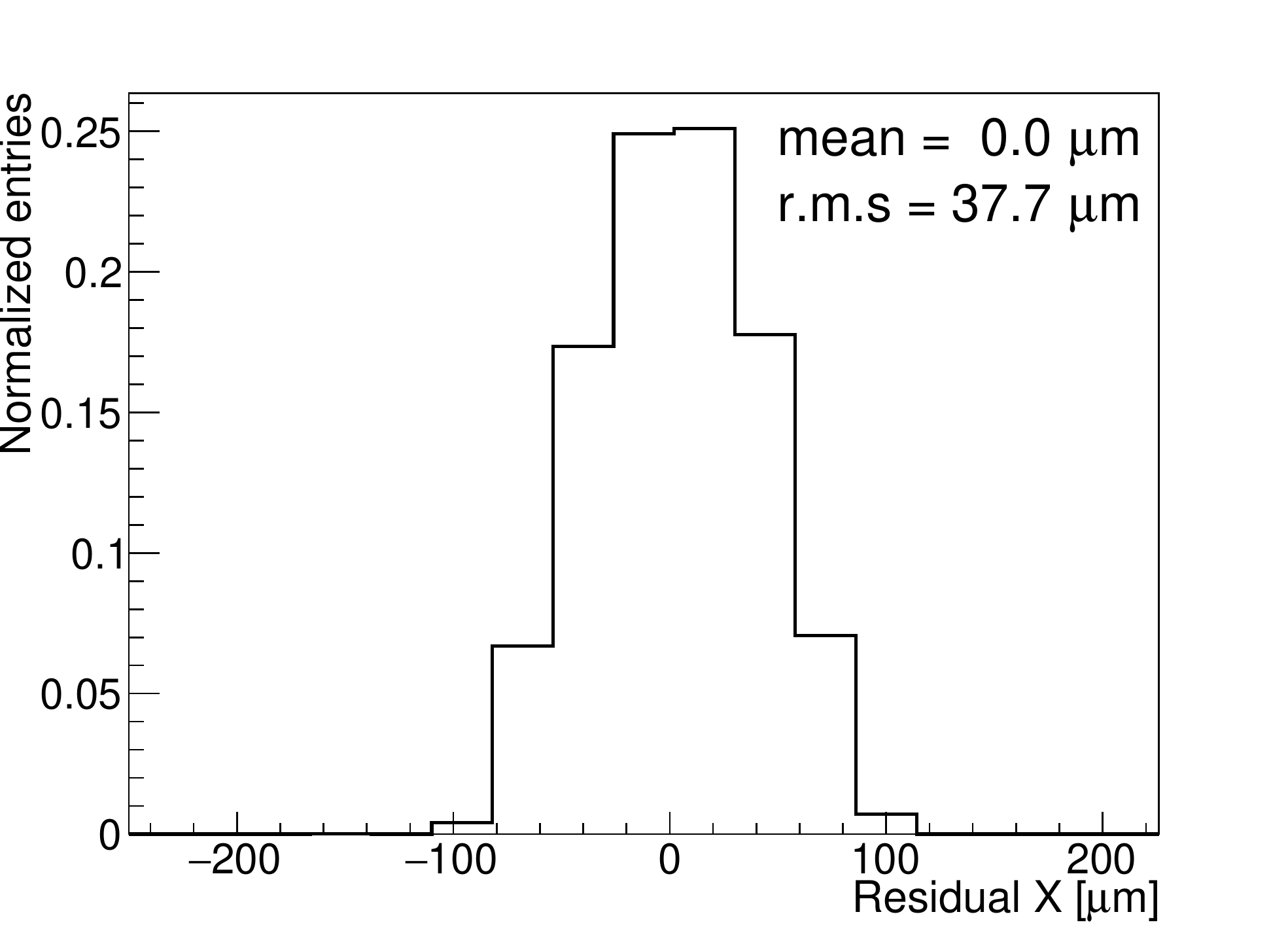}}\hfill
\subfigure[]{\includegraphics[width=.34\textwidth, ,trim=0cm 0cm 1.5cm 0.5cm, clip]{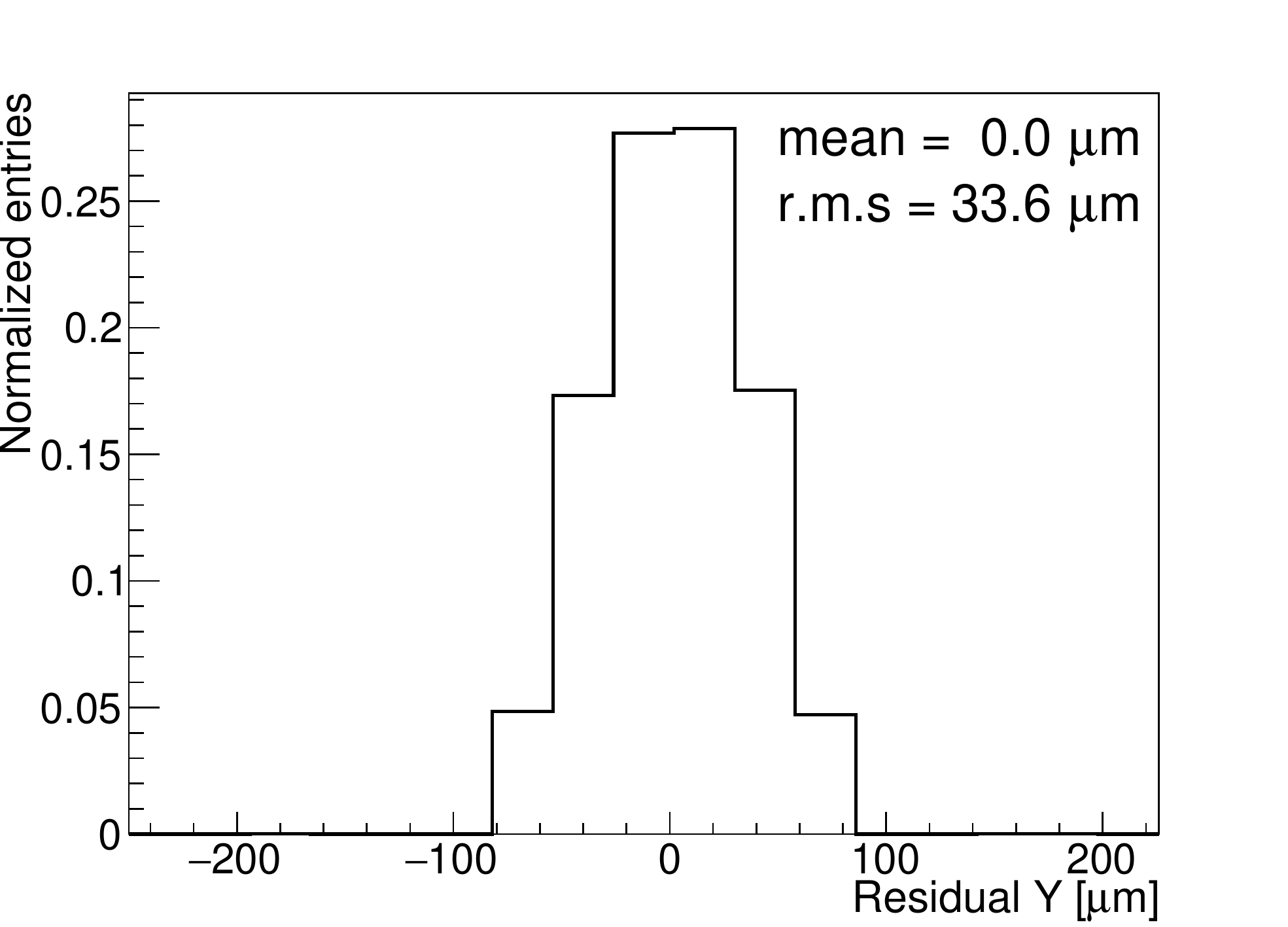}}\hfill
\caption{(a) Graphical depiction of the search area used to compute the hit efficiency, taken from \cite{CCPDv4-paper}. Normalised residual distributions of the $1\times10^{15}$  \Neq\  DUT along the $x$ (b) and $y$ (c) directions.}
\label{fig:hit-efficiency-def}
\end{figure}

The hit efficiency is calculated as the ratio between the number of clusters in the DUT that match reconstructed tracks from the beam telescope and the number of the good reconstructed tracks that are predicted to penetrate the DUT within its active area. The search area for tracks is an ellipse with the major and minor radii being $1.5$ macro-pixel pitches in $x$ and $y$ coordinates around the center of the cluster, i.e. 150 \micron\ and 187.5 \micron\, respectively (see also figure \ref{fig:hit-efficiency-def}a). Compared to the residuals of the neutron-irradiated CCPDv4 sample ($1\times10^{15}$  \Neq\ , figure \ref{fig:hit-efficiency-def}b and c), the size of the search area encompasses virtually the complete residual distributions. 

\begin{figure}[h]
\subfigure[]{\includegraphics[width=0.6\textwidth]{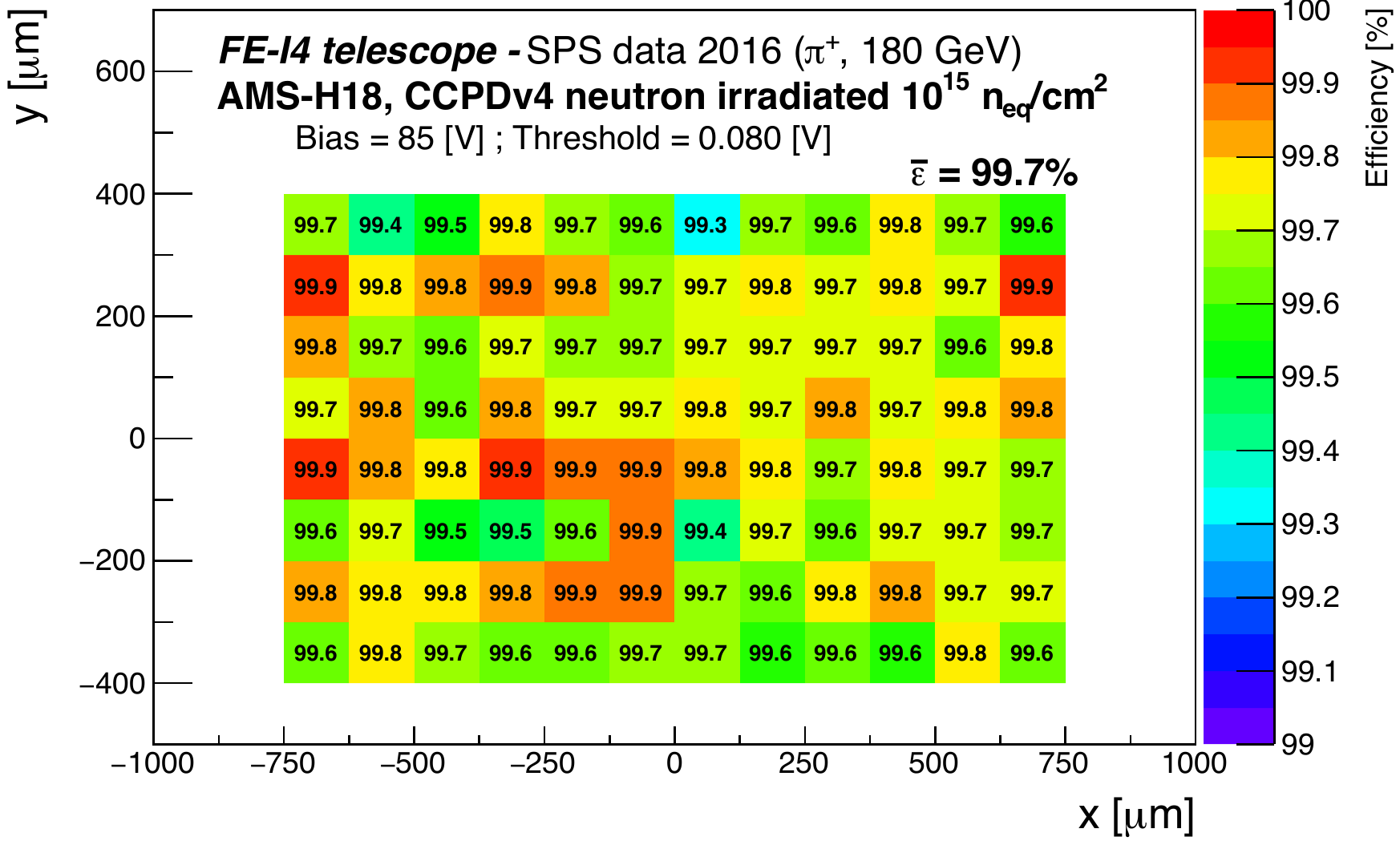}}
\subfigure[]{\includegraphics[width=0.4\textwidth, ,trim=0cm 0cm 0cm 0.5cm, clip]{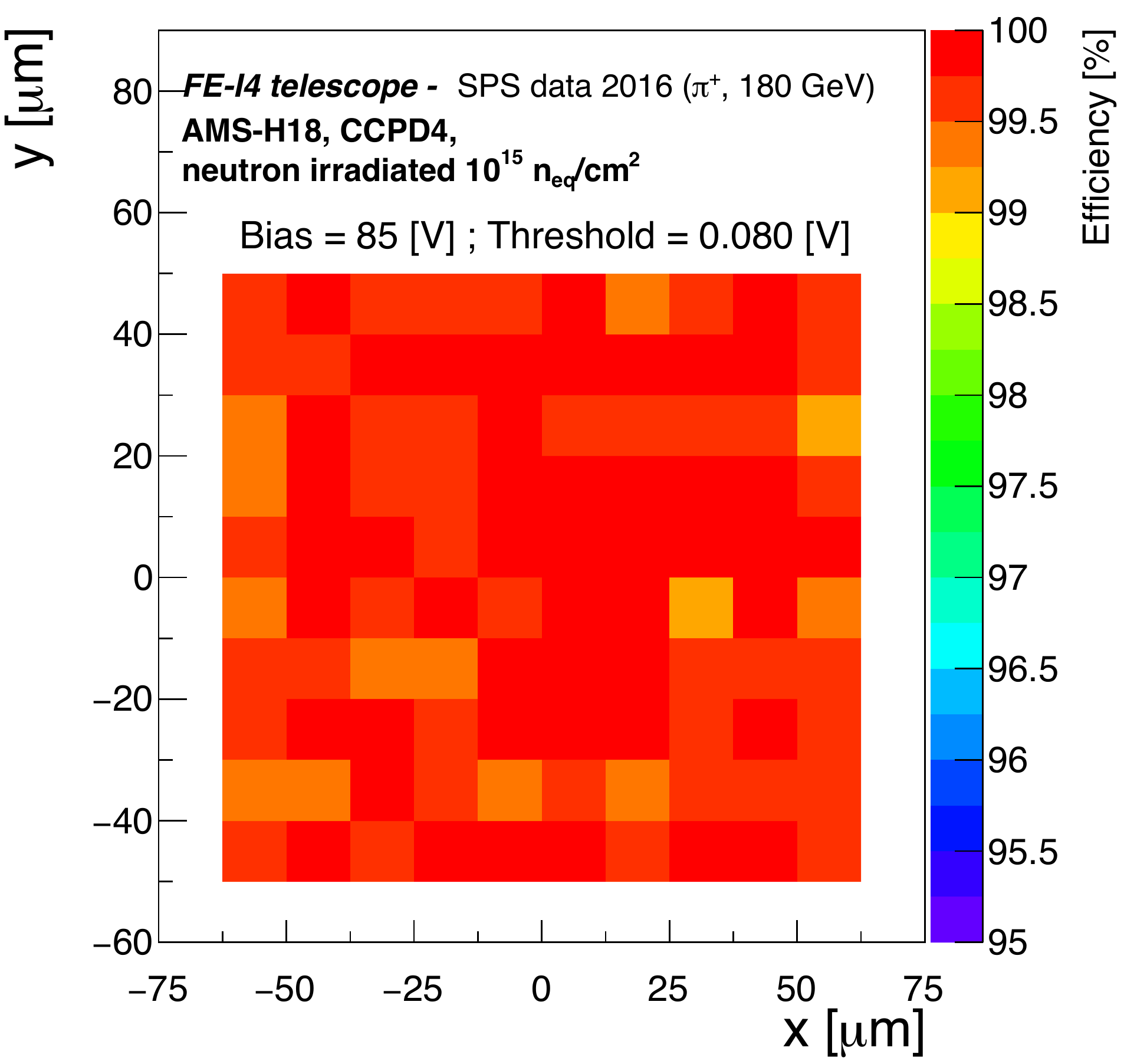}}
\caption{a) Hit efficiency map for CCPDv4 neutron-irradiated with  $1\times10^{15}$ \Neq; the colour scale is only ranging from 99\% to 100\%. b) Sub-pixel hit efficiency map overlaid from all central pixels. No significant efficiency loss is visible in any region of the pixel. The threshold of 80 mV is assumed to be equivalent to 690 e$^-$ according to \cite{CCPDv4-paper}.}
\label{caribou03_effmap}
\end{figure}

To avoid the excessive use of figures, the general properties will be introduced for one DUT ($1\times 10^{15}\Neq$, neutron-irradiated) and then comparison plots will be used to assess the radiation effects for different fluences.  The hit efficiency for the STime pixels in this sample at a bias voltage of 85~V and using a threshold of 80 mV is shown  in figure \ref{caribou03_effmap}a.

To avoid edge effects, the last 20 \micron\ at the edge of the outermost pixels have been excluded as this is where the telescope resolution ($1\sigma \approx 10\,\upmu$m, see \cite{telescopePaper}) leads to tracks being mis-reconstructed to be within the sensor active area while in reality the particle did not penetrate it. It can be seen that a very homogeneous and high efficiency of $99.7$ \% is reached across the whole matrix. By plotting the hit efficiency with sub-pixel resolution, it was found that there is no significant loss of hit efficiency in any region of the pixel (see figure \ref{caribou03_effmap}b).

\begin{figure}[!htb]
\begin{subfigmatrix}{2}
   \subfigure[Hit efficiency histogram for $1.3\times10^{14}$ \Neq.]{\includegraphics[width=.49\textwidth,keepaspectratio, trim=1cm 0cm 3cm 0.5cm, clip]{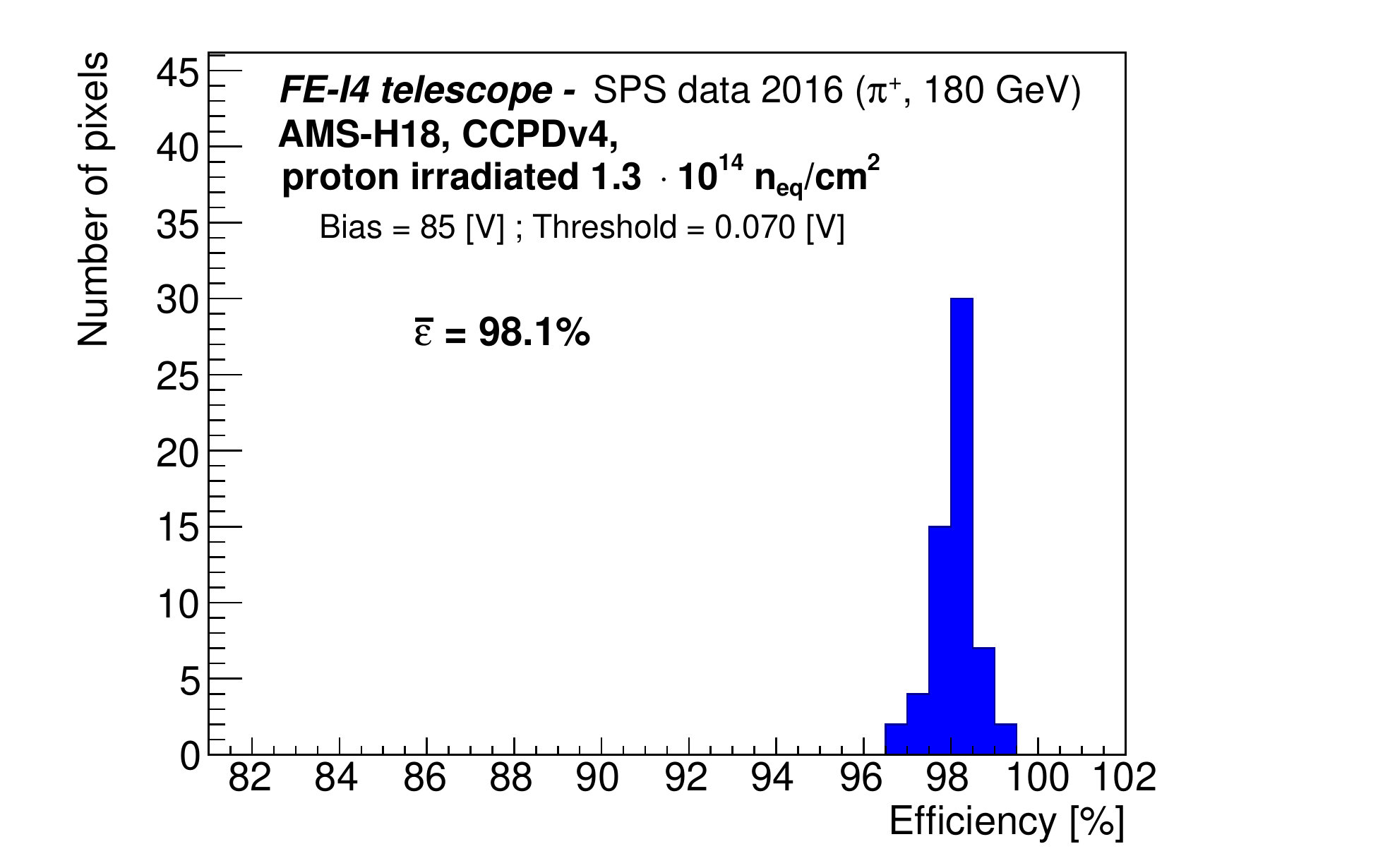}}
   \subfigure[Hit efficiency histogram for $5\times10^{14}$ \Neq.]{\includegraphics[width=.49\textwidth,keepaspectratio, trim=1cm 0cm 3cm 0.5cm, clip]{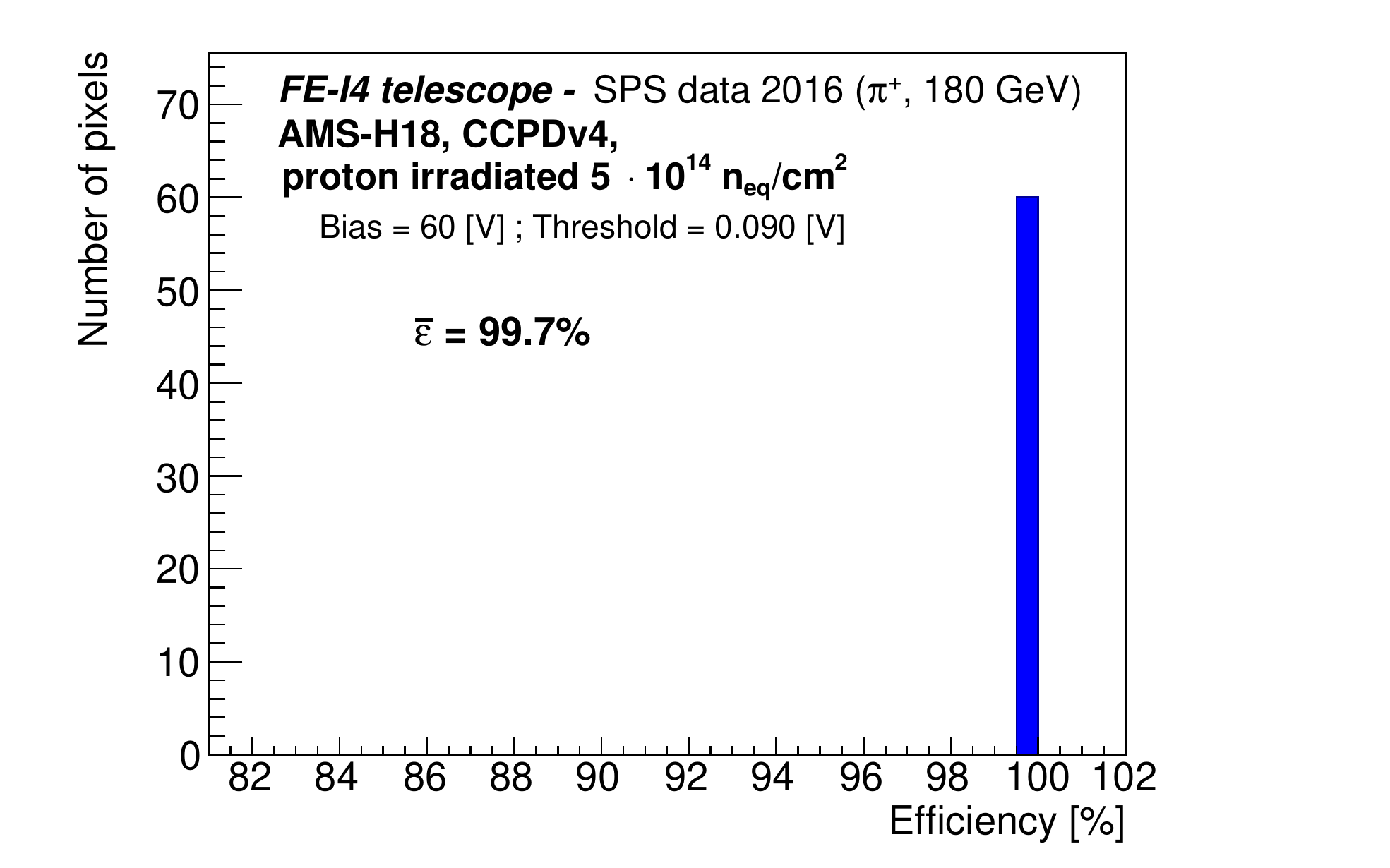}}
   \subfigure[Hit efficiency histogram for $1\times10^{15}$ \Neq.]{\includegraphics[width=.49\textwidth,keepaspectratio, trim=1cm 0cm 3cm 0.5cm, clip]{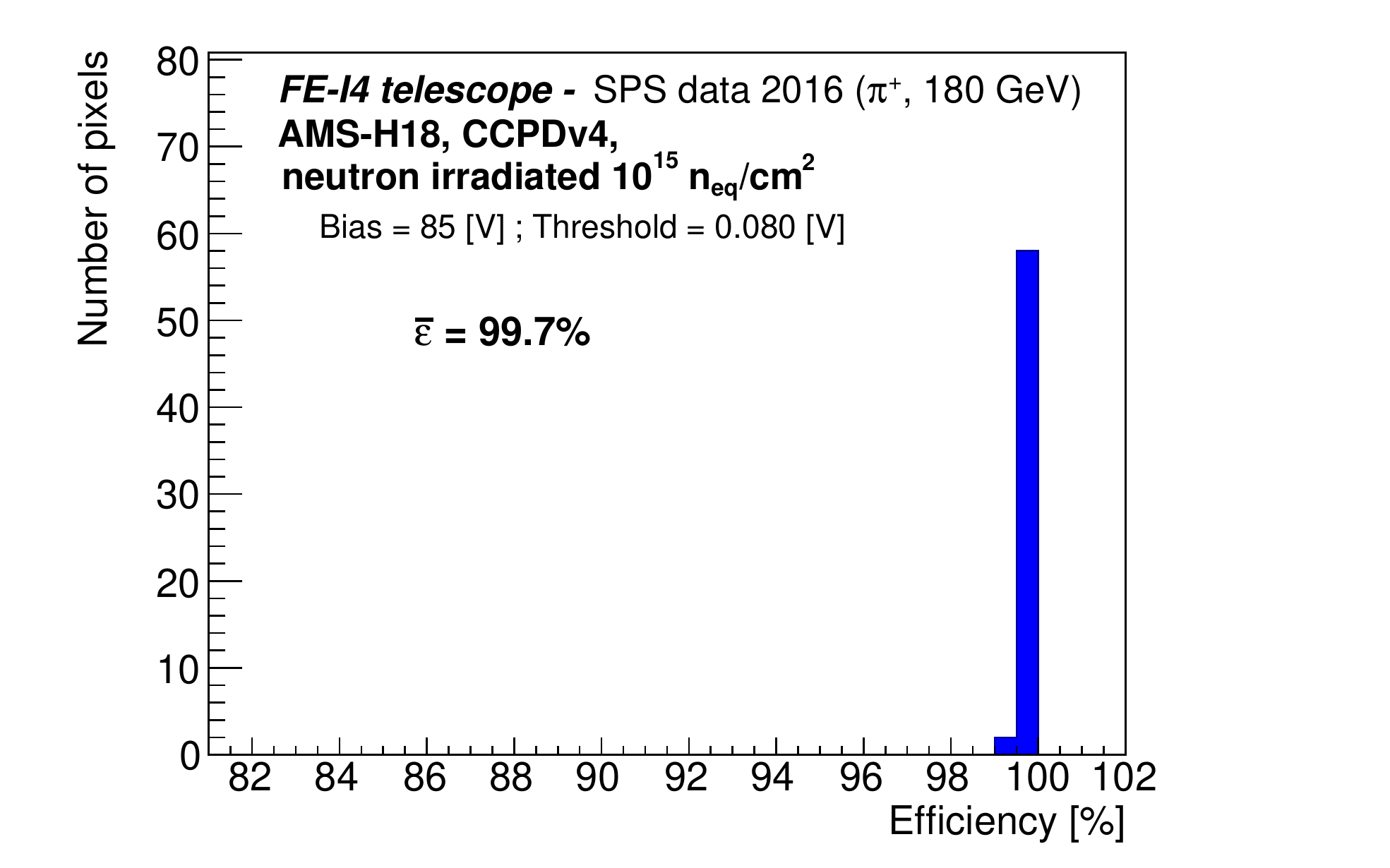}}
   \subfigure[Hit efficiency histogram for $5\times10^{15}$ \Neq.]{\includegraphics[width=.49\textwidth,keepaspectratio, trim=1cm 0cm 3cm 0.5cm, clip]{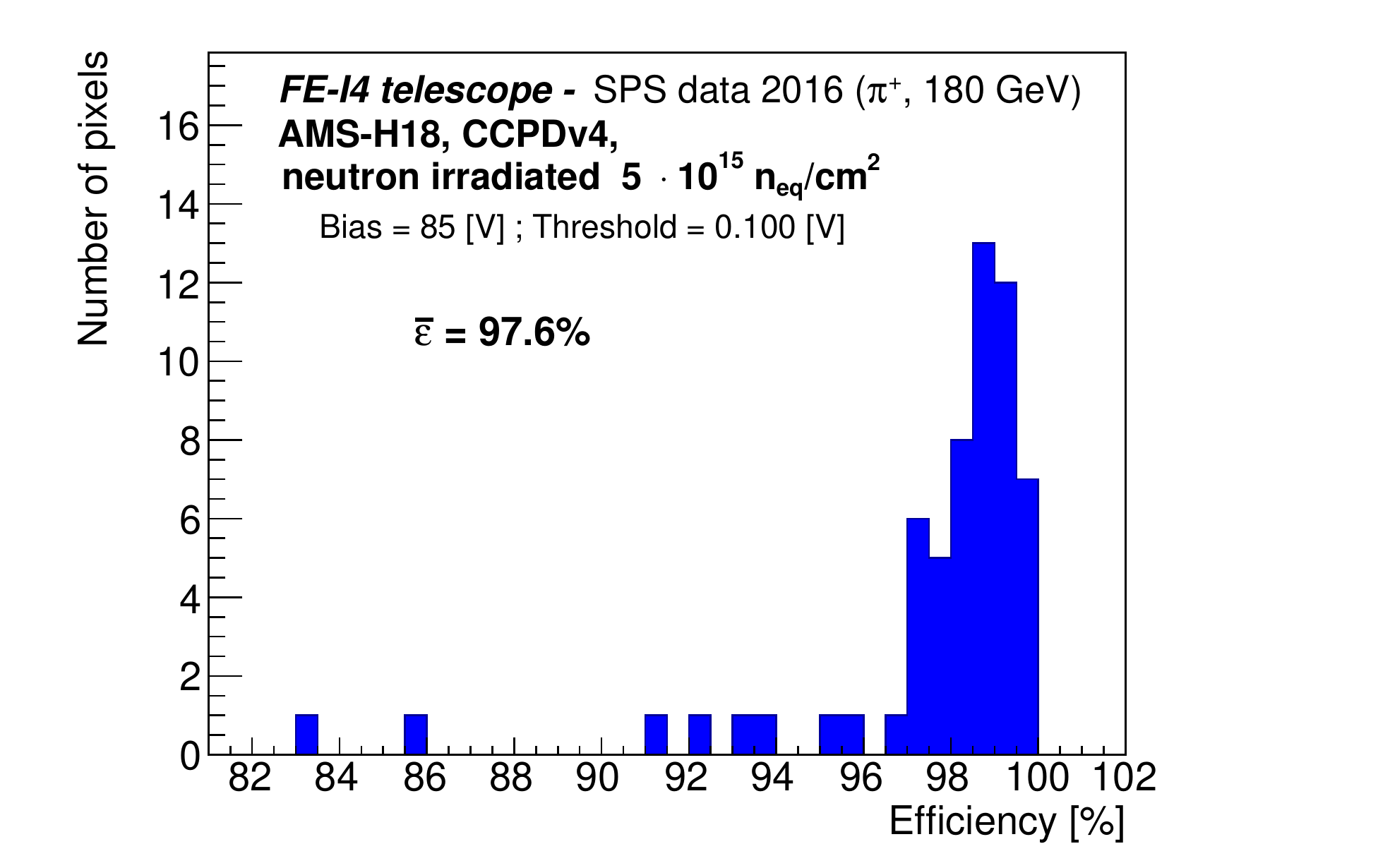}}
\end{subfigmatrix}
\caption{Hit efficiency histograms for different fluences of protons (upper row) and neutrons (lower row).The thresholds of 70, 80, 90 and 100 mV are assumed to be equivalent to 600, 690, 770 and 860 e$^-$, respectively, according to \cite{CCPDv4-paper}.}
\label{AllEff}
\end{figure}


A more effective representation of the the pixel hit efficiency distribution is by means of histograms, which allow to better judge outliers, for example due to defective sub-pixels. Figure \ref{AllEff} depicts hit efficiency histograms for samples irradiated with fluences of $1.3\times10^{14}$ , $5\times10^{14}$, $1\times10^{15}$  and $5\times10^{15}$ \Neq, which yield average hit efficiencies of $98.1$\%, $99.7$\%, $99.7$\% and $97.6$\%, respectively. These values are excellent and on a par with hit efficiencies of planar pixel sensors \cite{IBL-paper} -- for comparison, detector specs often require hit efficiencies of $>97$\% after irradiation. For the highest fluence, it can be seen that there are several outlier pixels, most probably originating from individual deteriorated circuits due to the rather high fluence of $5\times10^{15}$ \Neq. It is possible that in future improved designs, these could be recovered by extending the operating point of internal DACs, but in any case this would require further detailed studies at  circuitery level. It should be noted that for the  $5\times10^{14}$ \Neq\ sample, a working point at a bias voltage of only 60V was chosen because at HV$=80$V the sample showed a slightly higher noise occupancy, with the consequence of having a slightly decreased hit efficiency at 80V. 

\begin{figure}[b]
\begin{center}
\subfigure{\includegraphics[width=0.95\textwidth, trim=0cm 0cm 1.5cm 0.5cm, clip]{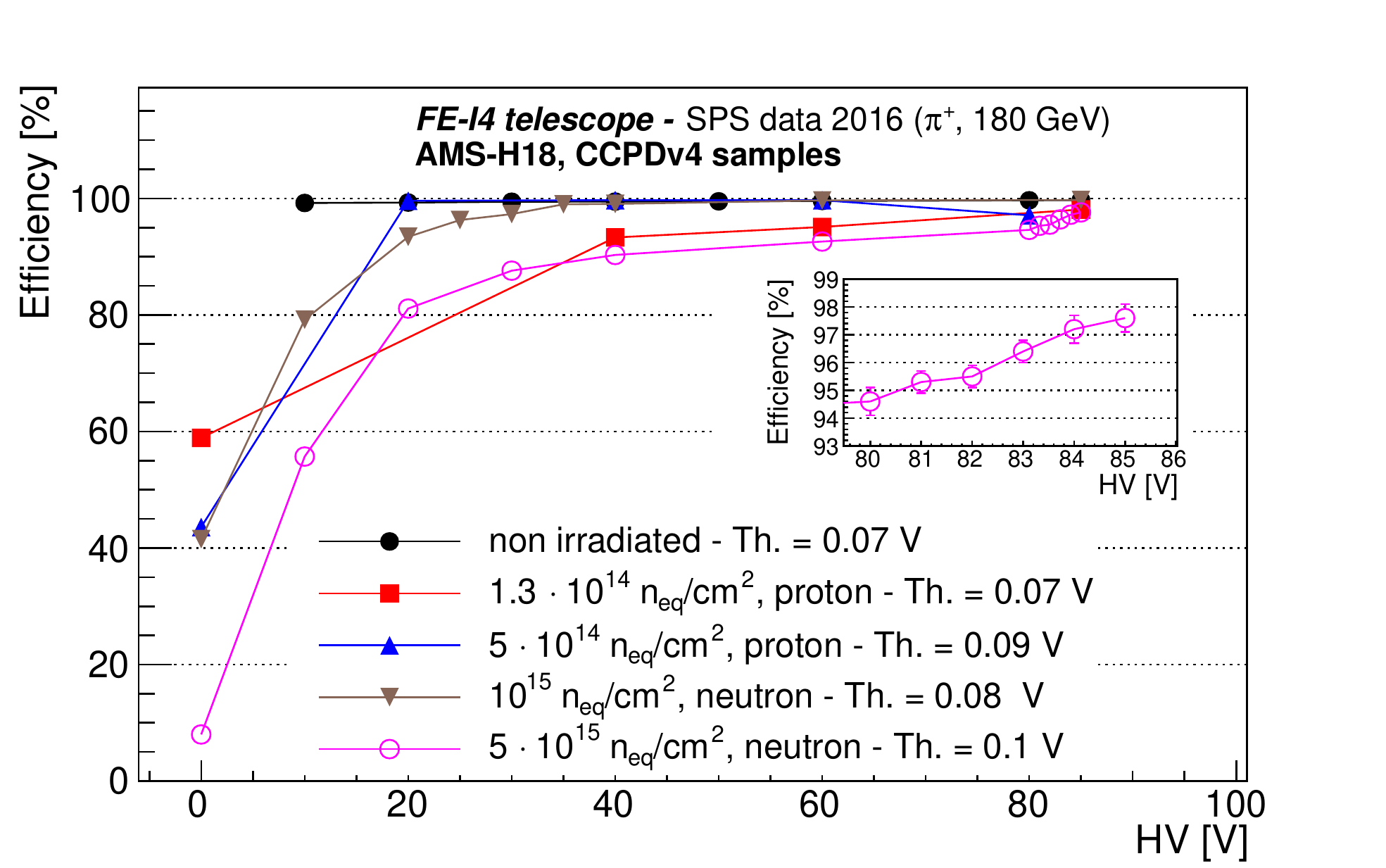}}
\caption{Average hit efficiency as a function of applied bias voltage. The inset shows the sudden increase in efficiency between 80 and 85 V, which could be attributed to charge multiplication. The thresholds of 70, 80, 90 and 100 mV are assumed to be equivalent to 600, 690, 770 and 860 e$^-$, respectively, according to \cite{CCPDv4-paper}.}
\label{HV_Scans}
\end{center}
\end{figure}

\begin{figure}[t]
\centering
\subfigure{\includegraphics[width=0.95\textwidth,trim=0cm 0cm 1.5cm 0.5cm, clip]{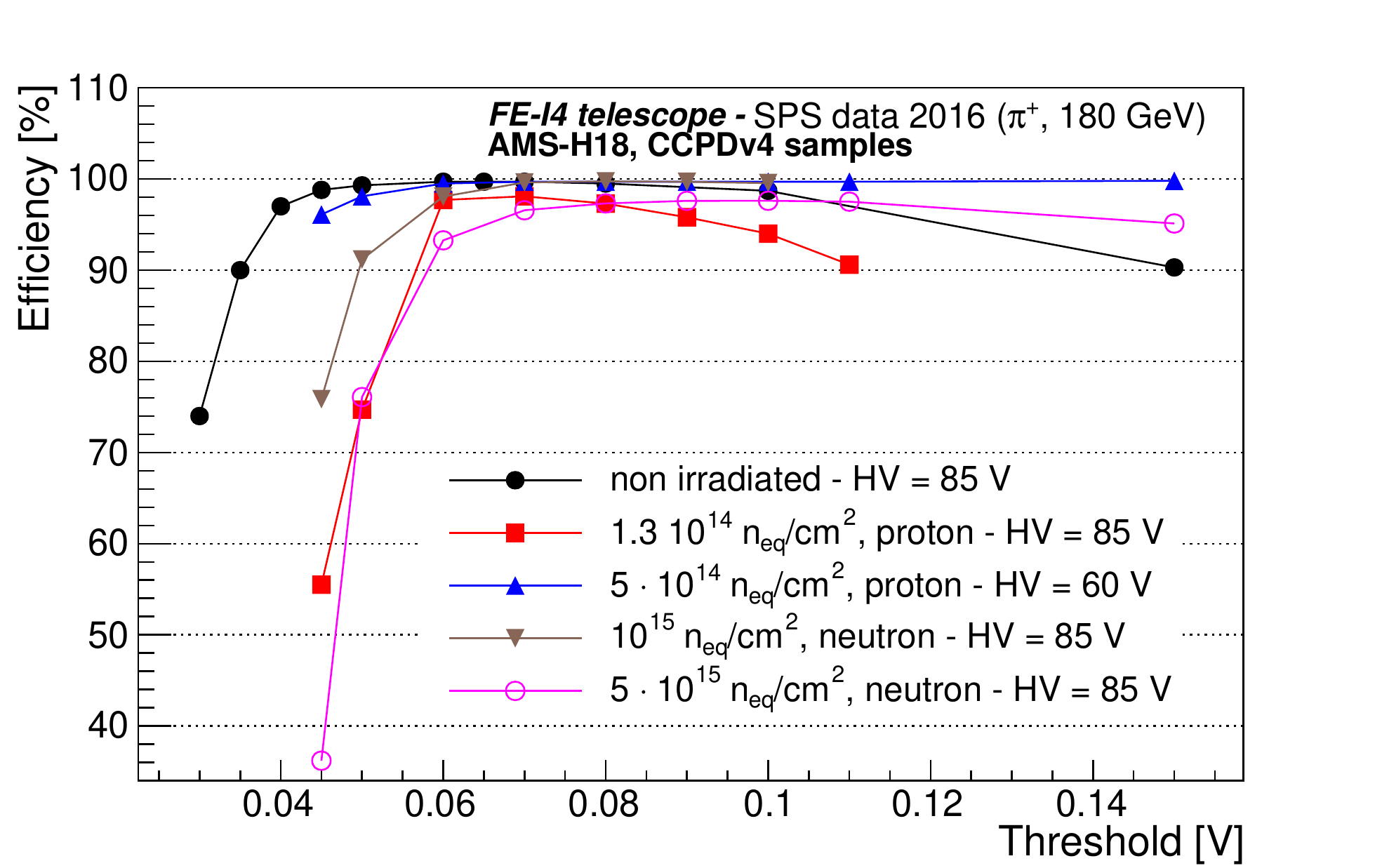}}
\caption{Average hit efficiency as a function of threshold voltage. For orientation: a thresholds of 60 mV is assumed to be equivalent to 520 e$^-$ according to \cite{CCPDv4-paper}.}
\label{TH_Scans}
\end{figure}

Bias-voltage and threshold scans were conducted to study their dependencies on the hit efficiency. The results are depicted in figure \ref{HV_Scans} and \ref{TH_Scans}. It can be seen that for the bias voltage, a plateau is reached after about 40~V, but in particular the very low and very high fluences profit from going to the highest possible bias voltages before the onset of breakdown, i.e.~about 85~V for this design. This is probably due to the fact that the depletion zone is thin compared to that of passive hybrid pixel sensors and its increase with bias voltage contributes significantly to the hit efficiency. For the middle fluences, the extension is already large enough thanks to the acceptor removal effect to not require the highest possible bias voltages. There are indications from other ams H18-based sensors that at high bias voltages, charge multiplication effects may increase the detectable charge \cite[p.~57 ff.]{Perrevoort}\cite[p.~64 ff.]{Hammerich}; this would be in line with the generally increased hit efficiency at bias voltages just below the breakdown. 

In highly irradiated (above $10^{15}$~\Neq) planar and 3D pixel sensors, it has been demonstrated that the high electric field combined with trapping due to radiation damage can lead to charge multiplication due to impact ionisation when sufficient high voltage is applied to the sensors \cite{Multi}. This fact generates an increase of the signal associated to the passage of a particle, which is correlated to the increase in noise and leakage current in the device. For the $5\times 10^{15}$~\Neq~  CCPD sample, a sudden increase in detection efficiency is observed between  80 and 85V, as shown in the insert of figure \ref{HV_Scans}. This measurement is correlated with a similar increase in leakage current of the device. Such a behavior was not expected and was not observed for the other devices irradiated with lower fluences and can be interpreted as evidence for the existence of the charge multiplication process in HV-CMOS devices, as previously observed in planar and 3D pixel sensors.

From figure \ref{TH_Scans}, three regions can be identified:
\begin{enumerate}
\item for very high thresholds there is a loss of events, and thus hit efficiency, due to low charge signals not being detected
\item for low threshold settings, the discriminators trigger on noise events and the dead time associated induces the measured inefficiency
\item in between there is a stable plateau, which is widest for the intermediate fluences, while for  $1.3\times10^{14}$ and $5\times10^{15}$ \Neq\  the lack of signal due to small depletion depth or increased trapping, respectively, reduces the span of the plateau region, i.e. of the operation point

\end{enumerate}
This is in line with earlier measurements \cite{Strip-CMOS} that saw an initial reduction in collected charge due to the loss of diffusing charge following the onset of trapping (at $\sim$ $10^{14}$ \Neq). This is followed by a strong increase in collected charge thanks to the increase of the depletion depth, which is the result of acceptor removal (at $\sim$ $10^{15}$ \Neq). At even higher fluences ($>5\times 10^{15}$ \Neq), the increasing trapping leads to reduction in collected charge again.

\subsection{Time Resolution}
For the HL-LHC ATLAS tracker, which will live in an environment with up to 200 pile-up events every 25\,ns, it is essential to be able to assign the different tracks to the correct bunch crossing (BC) time window. This requirement imposes the utilisation of fast shaping and fast readout, making classical MAPS\footnote{Monolithic Active Pixel Sensors} readout schemes, like rolling shutter, impossible. 

\begin{figure}[t]
\subfigure[]{\includegraphics[width=0.47\textwidth]{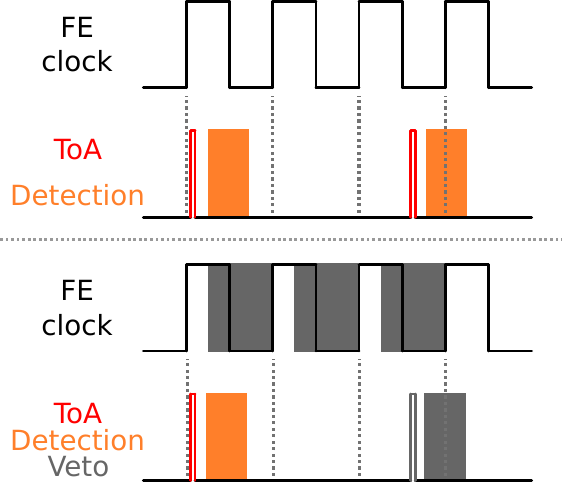}\label{fig:timing_cpv}}
\subfigure[]{\includegraphics[width=0.53\textwidth]{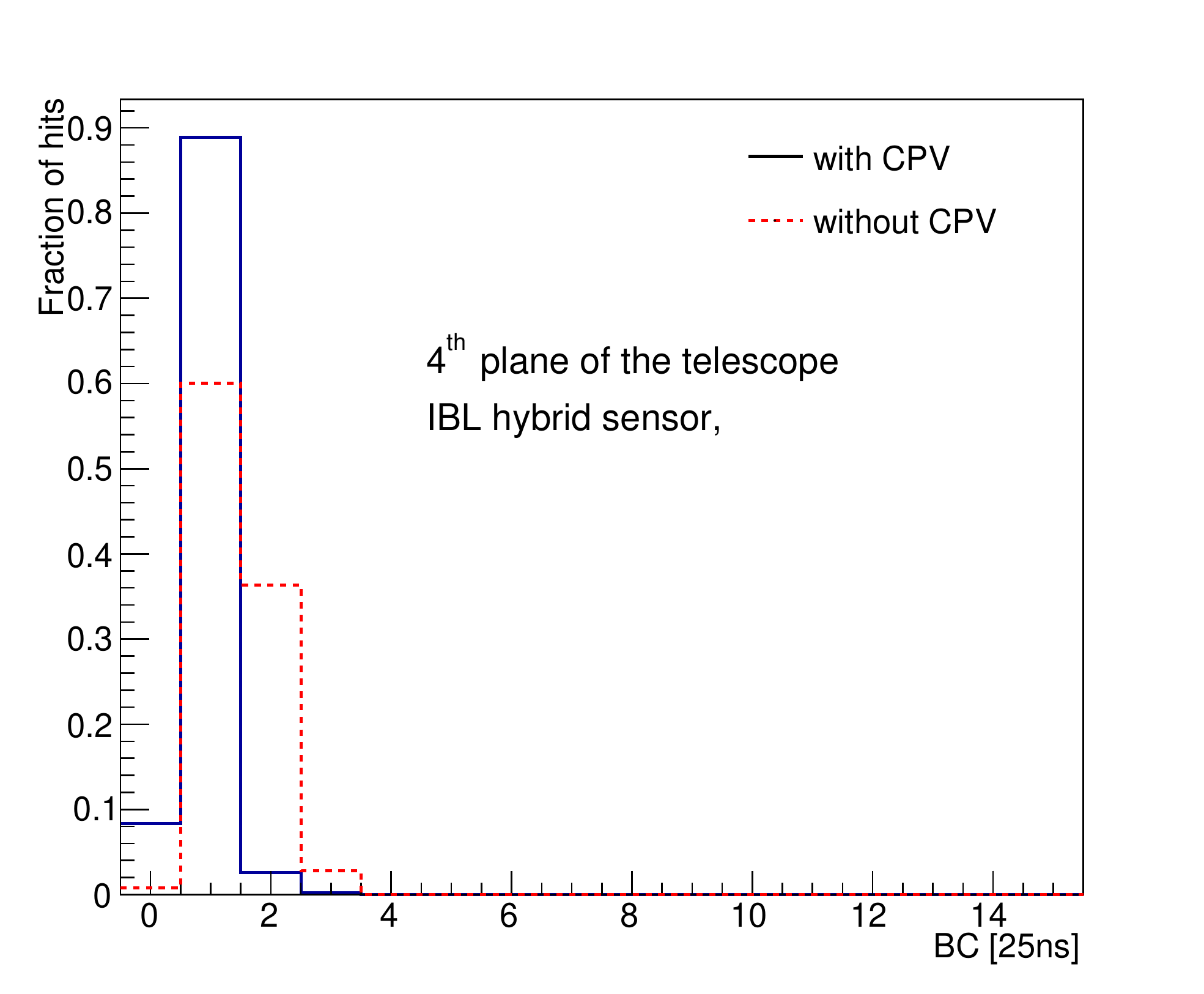}\label{fig:timing_lv1}}
\caption{a) Schematic of the Clock-Phase Veto. Triggers from particles arriving late in the FE-I4 clock cycle are discarded, emulating a bunched particle beam. b) Timing distribution of a planar sensor of the \emph{telescope plane} with and without Clock-Phase Veto.}
\end{figure}

In the CCPDv4 architecture, the trigger handling is done by the FE-I4 readout chip that samples the preamplifier signal every clock cycle of 25\,ns, which defines its time binning. However, due to time-walk of the FE-I4 preamplifier, particles depositing a small charge can sometimes be detected during the sampling interval next to the one of their arrival.
During operation in the ATLAS Detector, the FE-I4 clock is synchronised to the bunch crossings of the LHC beams, meaning that particles arrive in a narrow time window with respect to the rising edge of the FE clock.
By tuning the phase of the FE clock in a way that particles are detected mostly in one sampling interval, time-walk effects can be mostly mitigated in the LHC environment. 

\begin{figure}[!b]
\begin{center}
\subfigure{\includegraphics[width=0.65\textwidth]{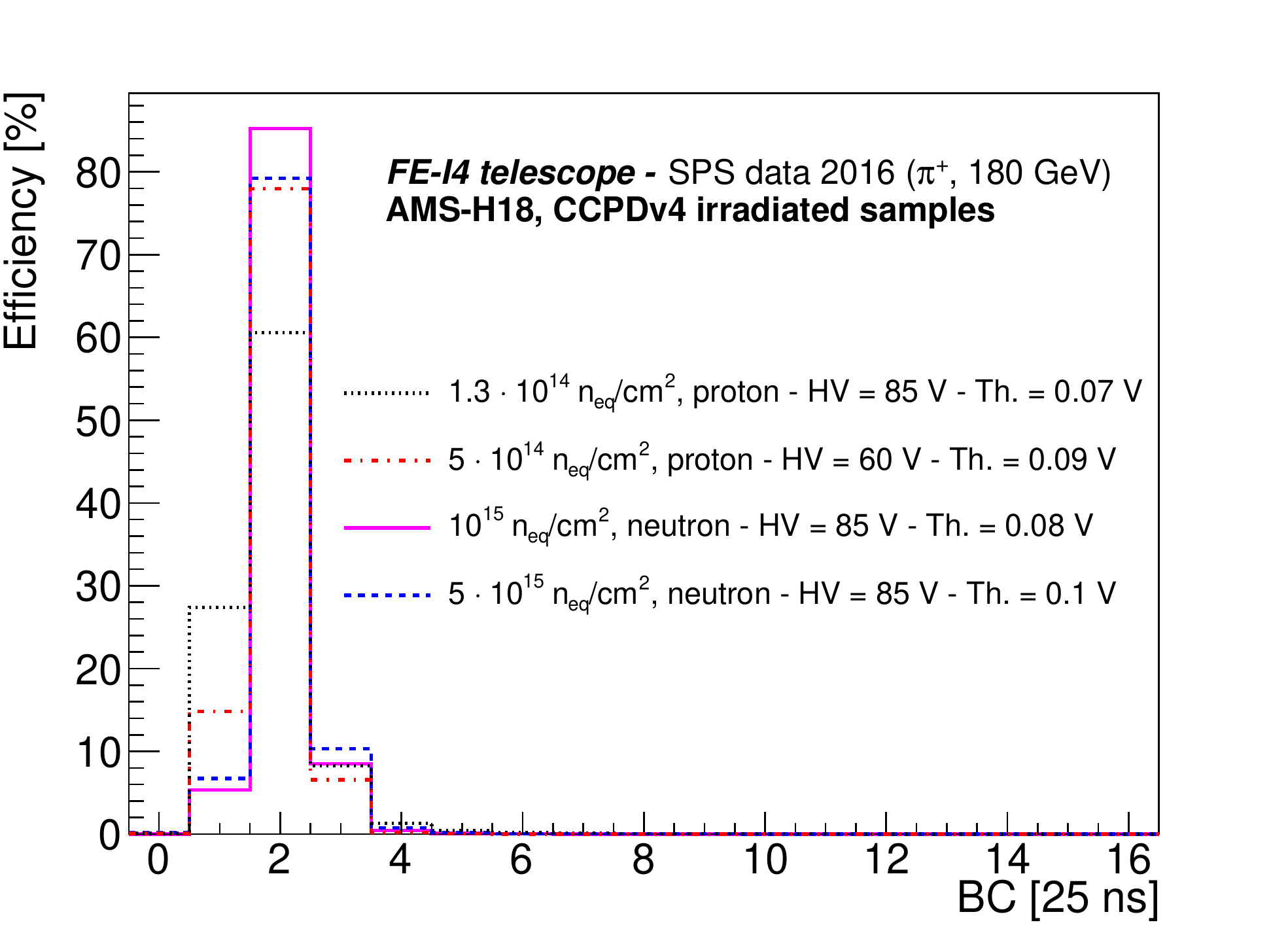}}
\caption{Timing distributions for the four irradiated CCPDv4 samples at the operational bias voltages. The thresholds of 70, 80, 90 and 100 mV are assumed to be equivalent to 600, 690, 770 and 860 e$^-$, respectively, according to \cite{CCPDv4-paper}.}
\label{Timing}
\end{center}
\end{figure}

The SPS H8 beam, however, has no pronounced timing structure over the $\sim$7\,s long spill.
Therefore, there is in general no coincidence of the time of arrival (ToA) of the particle and the FE clock, leading to an artificial smearing of the timing distribution. In order to mitigate this effect, a Clock-Phase Veto (CPV) as shown in figure \ref{fig:timing_cpv} was implemented.
Here, only particles that are detected in a  tunable interval of 6.4\,ns of the FE clock-phase can issue a trigger.
The effect of the veto for a telescope plane is depicted in figure \ref{fig:timing_lv1}.
All timing results in this paper were taken with the CPV; remaining timing delays are likely to be caused by either slow charge collection, e.g. in regions with low electric field strength, by slow rise time of the amplifiers in the CCPD circuits, or by time walk effect on the CCPD's discriminators.


Figure \ref{Timing} shows the timing distributions of the four irradiated samples. Clearly, most events are collected within one timing bin. It appears that a cumulative in-time efficiency of better than 95\% is achievable in 3 BCs for all fluences that have been investigated. This is comparable to the timing resolution of the FE-I4 telescope planes using passive 200 \micron\ thin n-in-n sensors (see figure \ref{fig:timing_lv1}), that have the same timing resolution as the baseline at least for the Pixel Endcap regions of the HL-LHC ATLAS Pixel Detector.

\begin{figure}[!htp]
\subfigure[]{\includegraphics[width=0.55\textwidth]{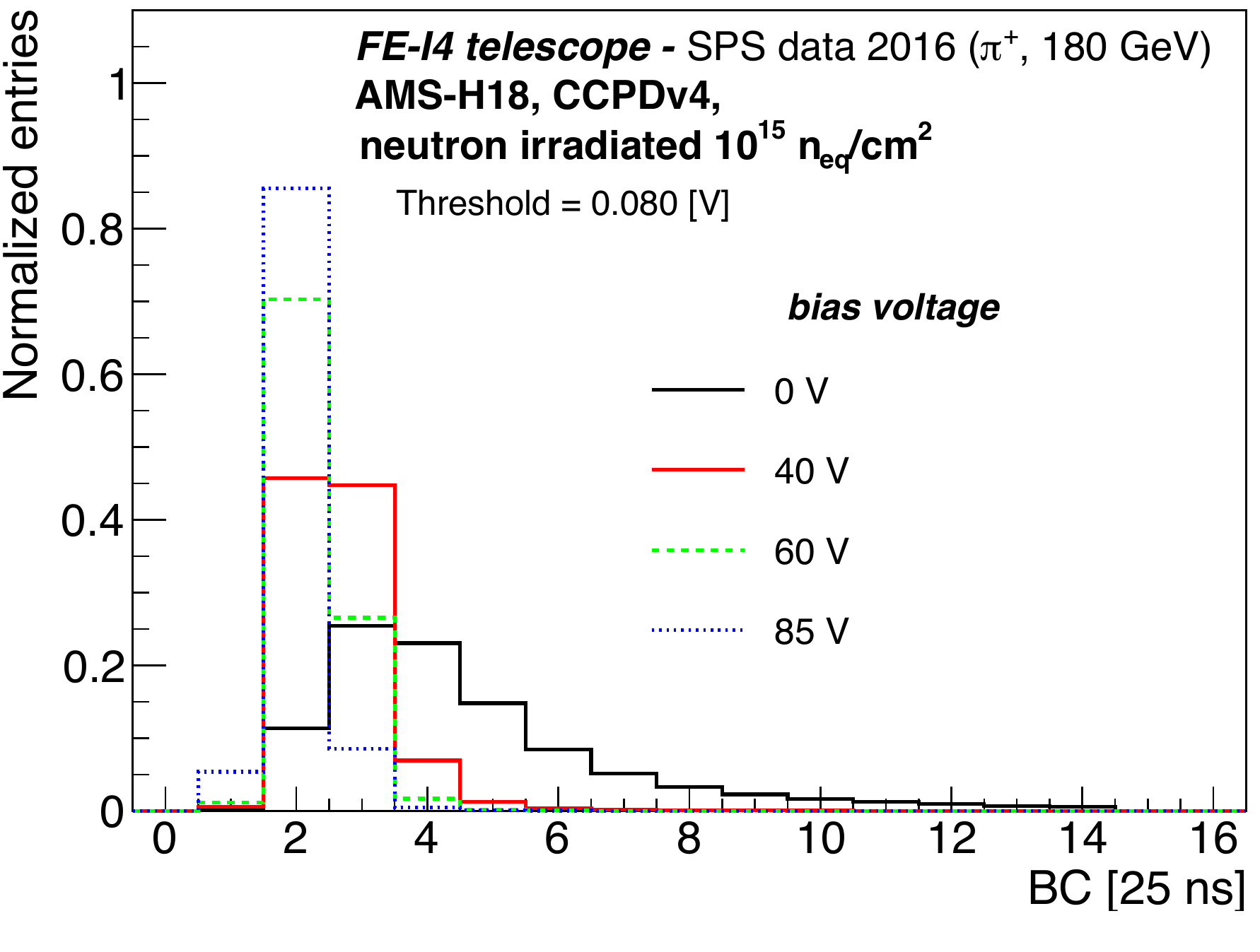}\label{Timing:voltage}}
\subfigure[]{\includegraphics[width=0.45\textwidth]{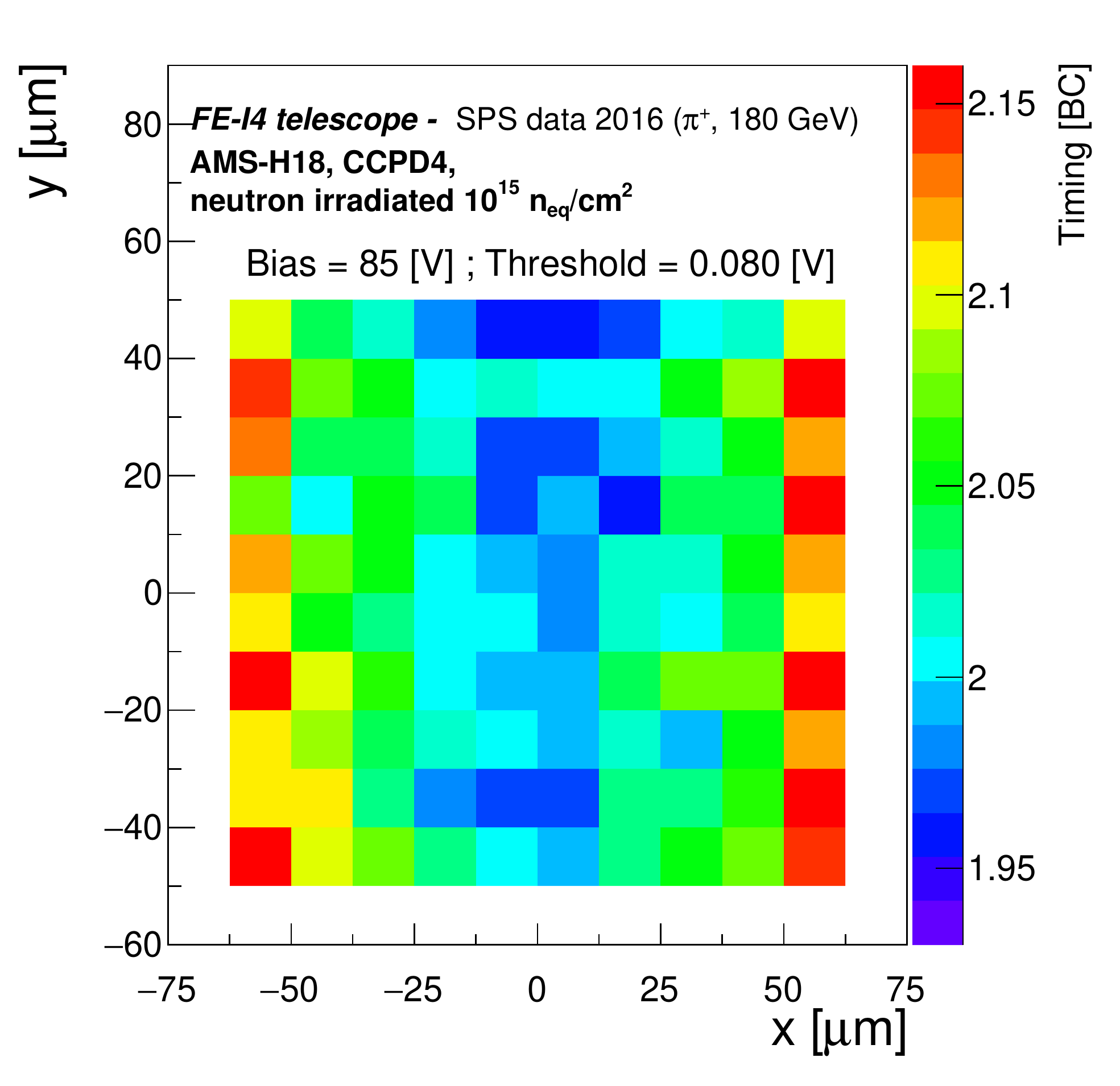}\label{Timing:InPixel}}
\caption{a) Timing for a CCPDv4 sensor for four values of the bias voltage. b) In-pixel timing map showing the mean timing of the event as depending on the track position inside the pixel. The threshold of 80 mV is assumed to be equivalent to 690 e$^-$ according to \cite{CCPDv4-paper}.}
\end{figure}

Some more insight can be gained when looking at the timing distribution for different applied bias voltages (figure \ref{Timing:voltage}) and at the average timing of an event depending on the track's in-pixel position (figure \ref{Timing:InPixel}). As expected, the events still registered without any applied bias voltage are slow and very broad in their timing. The time resolution gradually increases with the increasing electric field causing faster collection. The in-pixel mean timing map indicates that central events -- below or near the DNW -- are collected faster, while events at the edges of the pixel are on average slower. The CCPDv4 sensors are biased from the front side as there is no backside contact. TCAD simulations carried out for this study indicate that this may lead to regions of low electric field strength, supporting our measurements. Future tests to include a backside contact are planned.

\subsection{Cluster sizes}
Due to the small depletion depth, elimination of diffusion due to trapping and almost perpendicular beam incidence, very little charge diffusion is expected. Table~\ref{ClusterSize} shows this to be true, with the fraction of 2-hit clusters being reduced to the few \% level. This actually is an advantage, as average cluster sizes of 2 -- which are preferred thanks to their ability to yield better track resolution -- can be created by tilting the sensors with respect to the expected track direction. 

\begin{table}[ht]
\begin{center}
\caption{Cluster size fractions for different fluences.}
\vspace{2mm}
\begin{tabular}{|c||c|c|c|c|c|}
  \hline
  Fluence [\Neq] & 0 & $1.3\times10^{14}$ & $5\times10^{14}$ & $1\times10^{15}$  & $5\times10^{15}$  \\ \hline
  CS = 1 & 0.783 & 0.970 & 0.958 & 0.978 & 0.961 \\ \hline
  CS = 2 & 0.196 & 0.028 & 0.038 & 0.020 & 0.034 \\ \hline
  CS $\geq$ 3& 0.020 & 0.002 & 0.004 & 0.002 & 0.004 \\ \hline  
\end{tabular}\label{ClusterSize}
\end{center}
\end{table}

\section{Conclusions and Outlook}
Hybrid CCPDv4 prototype sensors produced in the ams H18 HV-CMOS process were glued to FE-I4 readout chips, irradiated up to $5\times10^{15}$~\Neq\ with reactor neutrons and 18 MeV protons and finally investigated in a testbeam experiment with high-energy pions at the SPS at CERN. Hit efficiencies up to $99.7\%$ at $1\times10^{15}$~\Neq\ were measured. Compared to unirradiated CCPD samples, they exhibit faster timing and less diffusion-based charge sharing. The improvement is attributed mainly to the increase in depletion depth as a consequence of reduced effective dopant concentration thanks to an acceptor removal effect in the low-resistivity ($10\,\Omega\cdot\mathrm{cm}$) p-type bulk.

The performance parameters after irradiation are impressive and close to being on a par with planar passive n-in-p pixel sensors. Further improvements can be expected by moderately increasing the base material resistivity. First sensors on such substrates (20, 80, 200 and 1000 $\Omega\,\cdot\,$cm) have already been produced in the ams H35 process. In addition to the hybrid CCPD approach, the small feature size of the H18 process allows the pursuit of fully monolithic designs, which would be an even larger benefit for the instrumentation of large areas. The first such prototypes are the MuPix8 and ATLASPix designs, which use the new aH18 process.

\acknowledgments
The authors gratefully acknowledge the support by the CERN PS and SPS instrumentation team and  would like to sincerely thank Prof.~Dr.~V.~Cindro and the team of the TRIGA reactor in Ljubljana for performing neutron irradiations for this publication. 
The research presented in this paper was supported by the SNSF grants  200021\_169015, 200020\_156083, 20FL20\_160474 and 200020\_163402. The irradiations at the TRIGA reactor in Ljubljana were supported by funding from the European Union's Horizon 2020 Research and Innovation programme under Grant Agreement no. 654168.

\end{document}